\documentclass[structabstract]{aa}  
\usepackage{graphicx}
\usepackage{amsmath}
\usepackage{txfonts}
\usepackage{natbib}
\usepackage[usenames]{color}
\begin{document}
\title{Solar velocity references from 3D HD photospheric models}

\author{
J. de la Cruz Rodr\'iguez\inst{1,2}, 
D. Kiselman\inst{1}, 
\and
M. Carlsson\inst{3,4}
}
\institute{Institute for Solar Physics, Royal Swedish Academy of Sciences, 
  AlbaNova University center, SE-106 91 Stockholm, Sweden
\and Department of Astronomy, Stockholm University, 
	AlbaNova University center, SE-106 91 Stockholm, Sweden
\and Institute of Theoretical Astrophysics, University of Oslo, 
  P.O. Box 1029 Blindern, N-0315 Oslo, Norway
\and
  Center of Mathematics for Applications, University of Oslo, P.O. Box 1053 Blindern, N-0316 Oslo, Norway
  }
\offprints{J.d.l.C.R.: \email{jaime@astro.su.se}}

\authorrunning{J. de la Cruz Rodr\'iguez et al.}
\titlerunning{Solar velocity references from 3D HD photospheric models}
\abstract 
    {The measurement of Doppler velocities in spectroscopic solar observations requires a reference for the local frame of rest. The rotational and radial velocities of the Earth and the rotation of the Sun introduce velocity offsets in the observations. Normally, good references for velocities are missing (e.g. telluric lines), especially in filter-based spectropolarimetric observations. }
    {We determine  an absolute reference for line-of-sight velocities measured from solar observations for any heliocentric angle, calibrating the convective line shift of spatially-averaged profiles on quiet sun from a 3D hydrodynamical simulation. This method works whenever there is quiet sun in the field-of-view, and it has the advantage of being relatively insensitive to uncertainties in the atomic data. }
    {We carry out radiative transfer computations in LTE for selected \ion{C}{i} and \ion{Fe}{i} lines, whereas the \ion{Ca}{ii} infrared lines are synthesized in non-LTE. Radiative transfer calculations are done with a modified version of \textsc{Multi}, using the snapshots of a non-magnetic 3D hydrodynamical simulation of  the photosphere.} 
    {The resulting synthetic profiles show the expected C-shaped bisector at disk center. The degree of asymmetry and the line shifts, however, show a clear dependence on the heliocentric angle and the properties of the lines.  The profiles at $\mu=1$ are compared with observed profiles to prove their reliability, and they are tested against errors induced by the LTE calculations, inaccuracies in the atomic data and  the 3D simulation.}
    {Theoretical quiet-sun profiles of lines commonly used by solar observers are provided to the community. Those can be used as absolute references for line-of-sight velocities. The limb effect is produced by the projection of the 3D atmosphere along the line of sight. Non-LTE effects on \ion{Fe}{i} lines are found to have a small impact on the convective shifts of the lines, reinforcing the usability of the LTE approximation in this case. We estimate the precision of the disk-center line shifts to be approximately 50~m~s$^{-1}$, but the off-center profiles remain to be tested against observations.}
\maketitle

\section{Introduction}\label{introduction}
The need for high spatial and spectral resolution in solar observations stimulates new advances in telescope and instrumentation technology, adaptive optics and image reconstruction techniques. However, the accuracy of the measurements is often limited by calibration issues. This is evident when using Doppler shifts of spectral lines to acquire line-of-sight velocities in the solar plasma, where the task of finding a local standard of rest is often problematic.

The current work is partly motivated by the installation of the CRisp Imaging Spectropolarimeter \citep[CRISP,][]{2008ApJ...689L..69S} at the Swedish 1-m Solar Telescope \citep[SST,][]{2003SPIE.4853..341S}. Instruments like this usually lack laboratory wavelength references, while the use of telluric lines has limitations as discussed below. The obvious (and commonly used) solution of letting the spectral line of interest -- averaged over a large enough area and long enough time -- define a local frame of rest is confounded by the fact that convective motions leave strong fingerprints on any spectral line formed in the solar photosphere. The statistical average of bright blueshifted and dark redshifted profiles formed in granules and intergranular lanes respectively, produces the  blueshifted C-shaped bisectors of photospheric lines as reviewed by \citet{1982ARA&A..20...61D}.

Below, we discuss some methods that have been used to define a velocity reference for solar observations.
\begin{itemize}
	\item \emph{Sunspot umbrae} are common standards \citep[e.g.,][]{1977ApJ...213..900B,2008ApJ...689L..69S,2010ApJ...713.1282O}, but this only works  when a suitable sunspot is in the field-of-view and relies on the assumption that the umbra is at rest because of the convective motions being suppressed by the strong magnetic fields. Furthermore, the line must be measurable in the umbra and wavelength shifts which are induced by blends from lines uniquely formed in the umbra (e.g. molecules), must be accounted for. Finally, observations of umbrae are always plagued by stray light from the much brighter quiet photosphere.

	\item \emph{Telluric lines} can be used to define a laboratory-frame reference, which can then be converted to the Sun. \citet{1997ApJ...474..810M} and \citet{2008ApJ...676..698B} used telluric lines to derive the wavelength scale for their observations, then converted to an absolute wavelength scale on the Sun given the ephemeris constants, time of the observations, solar rotation and the laboratory wavelengths for all observed spectral lines.

	\item \emph{A spectral atlas}, most commonly  the atlas acquired with the Fourier Transform Spectrometer at the McMath-Pierce Telescope (hereafter called the FTS atlas) of \citet{fts-atlas}, can be used to calibrate observations. This was done by \citet{2007ApJ...655..615L}, who used the laboratory wavelength to define the reference for velocities, assuming an absence of systematic errors in the FTS atlas itself and correcting for the gravitational redshift (633~$\mbox{m}\ \mbox{s}^{-1}$).  This method only works at disk center where the atlas was recorded.
	\item  \emph{Numerical models} can be used to predict convective line shifts and thus provide a velocity reference. A two-component model of the solar photosphere has been derived by \citet{2002A&A...385.1056B} from the inversion of \ion{Fe}{i} spectral lines. Since it does not contain any horizontal velocities, it then is in principle limited to calibrating disk-center observations, for which it has been used by \citet{2004A&A...415..717T} and \citet{2009A&A...508.1453F}. However, \citet{2004A&A...427..319B} used it with the empirical results of \citet{1988A&AS...72..473B} to also estimate lineshifts off solar center. \citet{2007ApJ...655..615L} employ a 3D numerical granulation model to compute the convective blueshift of the \ion{C}{i}~5380~\AA\ line and calibrate their observations. This approach was necessary  because the laboratory wavelength of the \ion{C}{i} line is not known with enough precision to use the atlas calibration that is mentioned above.
\end{itemize}

We note that telluric lines cannot always be observed with the solar diagnostic lines and that when this is the case, observing them can be expensive. Observations with tunable filters require high cadence because of seeing variability and solar evolution. Any time spent recording telluric lines thus decreases the spatial resolution and the signal-to-noise of the science data. This would also apply to any laboratory spectral-line source used for calibration. Furthermore, approaches using telluric lines (or other calibration lines) and spectral atlases all require very accurate atomic data to define the zero velocity point and thus convert the wavelength scale into velocity scale. Such data are unfortunately not available for many of the lines that are extensively used in solar physics as is apparent in the work of \citet{2007ApJ...655..615L} described above. Their approach is proposed in this study: using realistic 3D numerical simulations of the solar photosphere to compute spatially-averaged line profiles. As the radiative transfer computation is carried out in the local frame of rest of the Sun and the adopted atomic data is known, the velocity shifts of the spatially-averaged profile can be computed accurately for different heliocentric angles, relative to the assumed center of the line. This method relies on the degree of realism of the 3D simulation and requires a large statistical sample of spectra to compute the spatially-averaged profile. \\
The aim of this work is to provide the community with a set of theoretical  spatially-averaged quiet-Sun profiles of lines commonly used in solar physics, that are computed for a range of heliocentric angles. 

In Sect.~\ref{llist} the atomic data and radiative transfer details are discussed. The 3D HD simulation is described in Sect.~\ref{hydromod}. In Sect.~\ref{res} the results are presented together with an error analysis and a real filter-based observation is calibrated using the new data. Section~\ref{edata} describes the electronic data made available and Sect.~\ref{conclusions} summarizes the main conclusions of this work.
\section{Line list and radiative transfer}\label{llist}
The calculations were carried out with a selection of lines (Table~\ref{input}) that are extensively used in solar physics, including  lines that can currently be observed with CRISP at the SST.
A modified version of the radiative transfer code \textsc{Multi} \citep{1986UppOR..33.....C} was used to compute synthetic spectra from the simulation snapshots. The \ion{Fe}{i} and \ion{C}{i} lines were calculated assuming local thermodynamical equilibrium  (LTE), which is a reasonable assumption for the wings of most photospheric \ion{Fe}{i} lines \citep{1999A&A...350.1018G,2000A&A...359..729A}. \citet{2006A&A...458..899F} found that non-LTE effects for \ion{C}{i} lines are expected to be small with the effect on derived abundance being $<0.05$ dex. Possible errors produced by non-LTE effects will be further discussed in Section \ref{non-LTEe}. 

The \ion{Ca}{ii} lines were computed in 1.5D non-LTE -- where the non-LTE problem is solved in 1D --  using the same model atom as \citet{2009ApJ...694L.128L}, which consists of 5 bound levels plus a continuum.

Atomic data were obtained from the VALD-2 database \citep{2000BaltA...9..590K}. The treatment of van der Waals broadening follows that of \cite{1998PASA...15..336B} and the required cross sections were taken from \cite{2000yCat..41420467B}, except for the \ion{C}{i} line provided by Paul Barklem (private comunication). The effective quantum numbers $n*$ for this transition are $1.95 \rightarrow 3.27$. The treatment becomes less reliable when $n*>3.0$ but it is more accurate than the commonly used \cite{1955QB461.U55......} formulation.
\begin{table}[]
\begin{minipage}[]{\hsize}
 \caption{Atomic data, sorted by element in increasing wavelength.}
 \label{input}
 \centering
\begin{tabular}{c|ccccc}
  \hline
  \hline
  $\lambda_0$ (\AA)& $E_c$ & log $gf$ & $\sigma^{(1)}$ & $\alpha^{(2)}$ & log $\epsilon^{(3)}$ \\
  \hline
  \ion{C}{i} 5380.3370   & 7.6850& -1.824 & 1238 & 0.229 & 8.53 \\
  \ion{Fe}{i} 5250.2089  & 0.1210& -4.938 & 207 & 0.253 & 7.44\\
  \ion{Fe}{i} 5250.6460  & 2.1980& -2.181 & 344 & 0.268 & 7.71\\
  \ion{Fe}{i} 5576.0888  & 3.4300& -1.000 & 854 & 0.232 & 7.69\\
  \ion{Fe}{i} 6082.7106  & 2.2230& -3.573 & 306 & 0.271 & 7.45\\
  \ion{Fe}{i} 6301.5012  & 3.6540& -0.718 & 834 & 0.243 & 7.55\\
  \ion{Fe}{i} 6302.4940  & 3.6860& -1.203 & 850 & 0.239 & 7.50\\
  \ion{Fe}{i} 7090.3835  & 4.2300& -1.210 & 934 & 0.248 & 7.63\\
  \ion{Ca}{ii} 8498.0134 &  1.6920 &   -0.6391 & 291 & 0.275 & 6.31\\ 
  \ion{Ca}{ii} 8542.0532 &  1.7000 &   -0.4629 & 291 & 0.275 & 6.31\\
  \ion{Ca}{ii} 8662.1605 &  1.6920 &   -0.7231 & 291 & 0.275 & 6.31\\
  \hline 
\end{tabular}
\tablefoot{$(1)$ the cross-section in units of the Bohr radius, $(2)$ dimensionless parameter used with $(1)$ in the computation of collisional broadening \citep{1998PASA...15..336B}. $(3)$ the fitted value of the abundance for each line.}
\end{minipage}
\end{table}

To achieve the best possible fit between the synthetic line profile at $\mu=1$ and the FTS atlas, the elemental abundance is individually adjusted for each line. The resulting abundance values (last column in Table~\ref{input}) should not be regarded as such. Instead, they act as free parameters that compensate for errors from atomic data, the LTE approximation, and the models themselves. The hope is that by fitting the profile the correct layer of the model with the appropriate velocity field is being sampled. Since non-LTE effects could be expected to affect the cores of strong lines, only the wings of each line were used for the fit of the abundance; however, the results change marginally when the full profile is used to fit.

The results for the \ion{Ca}{ii} infrared triplet lines must be approached with extra caution since the lines are very strong and the 3D simulation does not extend all the way up to the chromosphere. Therefore the calculated core of the line cannot be compared to real observations, but the photospheric origin of the wings suggests the usability of the models with these lines, as discussed by \citet{1974SoPh...37..145S}. 

We have included blends affecting the \ion{Fe}{i} and \ion{C}{i} lines. The atomic parameters for these were taken from the VALD-2 database \citep{2000BaltA...9..590K}. The blends are weak, but have a small effect on the upper part of some bisectors. The \ion{Ca}{ii} lines have some strong blends in the wings which greatly affect the bisectors, so these blends were not present in the present \ion{Ca}{ii} calculations.

\section{3D hydrodynamical photospheric models}\label{hydromod}
The 3D hydrodynamical simulation consists of 90 snapshots and covers around 45 minutes of solar time \citep[Trampedach et al., in prep.;][]{2009A&A...507..417P}.  The simulation does not include magnetic fields. The radiative energy transport was computed with an improved multi-group opacity binning of 12 bins, when compared with the 3D models of \citet{2000A&A...359..729A}. The original simulation of $240\times 240\times 240$ grid-points covering a region of $6\times 6 \times 4$ Mm was interpolated to a finer vertical depth grid in the radiative zone and sampled to a coarser grid in the horizontal plane for radiative transfer calculations. The resulting snapshots had a size of $50\times50\times82$ grid points representing a 6~Mm square. Our tests show that the effect of such a resampling in the horizontal plane has a negligible effect on the spatially-averaged profiles of the lines.

\begin{figure}[]
      \centering
      \resizebox{\hsize}{!}{\includegraphics[]{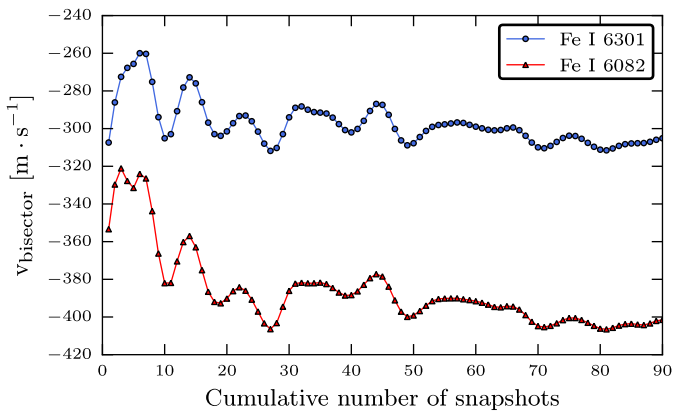}}
        \caption{
		Mean bisector velocity as a function of the number of snapshots used to compute the spatially-averaged spectrum. 
       }
        \label{osci}
    \end{figure}

The presence of oscillations in the simulation broadens the time-averaged profile compared to the individual snapshots. The oscillations have a quasi-periodic behavior, so at least one period should be included in the time average in order to avoid systematic shifts of the average profiles. Figure~\ref{osci} shows the average velocity (integrated along the bisector) of the time averaged profile as a function of the number of snapshots used in the time series. The limited sample of granules and intergranular lanes enclosed in a $6$ Mm box reinforces the need to use more than one period in order to converge to a stable average for the line shift. Therefore, the whole time series was used for computing the average line profiles.

To produce profiles off disk center, the properties of the 3D snapshots are slanted assuming periodic boundary conditions, and the velocity vector is projected into the new line-of-sight. The height scale is also modified to match the new geometry of the 3D model.

\section{Results} \label{res}
\begin{figure*}[]
     	\resizebox{\hsize}{!}{\includegraphics[]{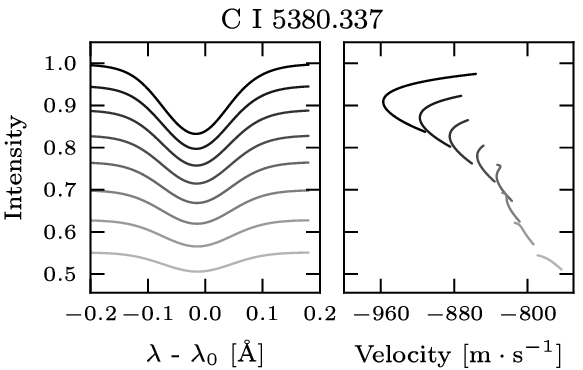}\includegraphics[]{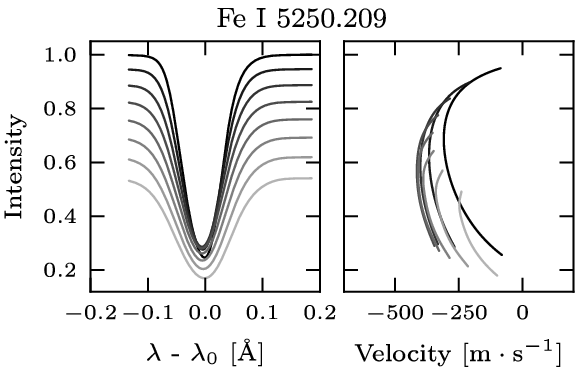}\includegraphics[]{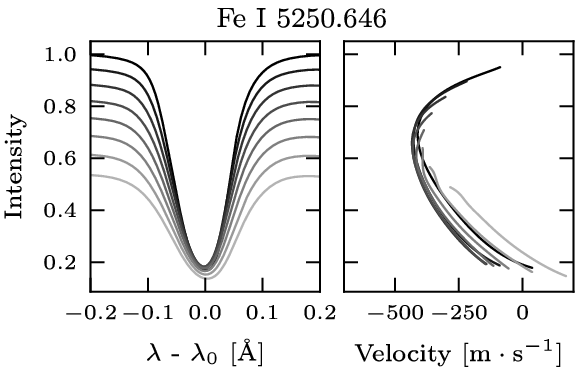}}
     	 \resizebox{\hsize}{!}{\includegraphics[]{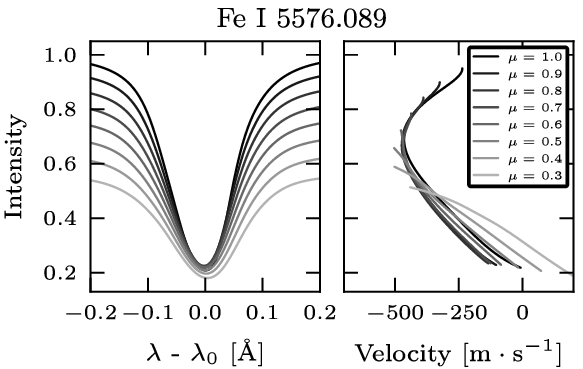}\includegraphics[]{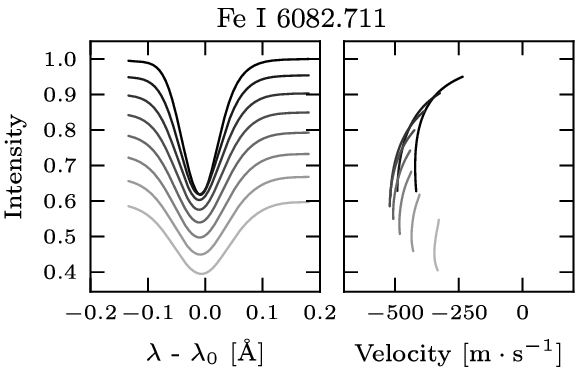}\includegraphics[]{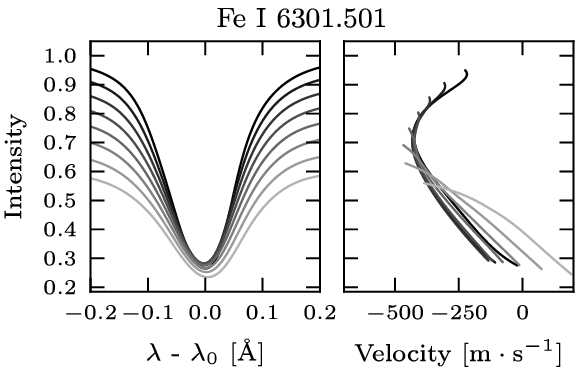}}
     	 \resizebox{\hsize}{!}{\includegraphics[]{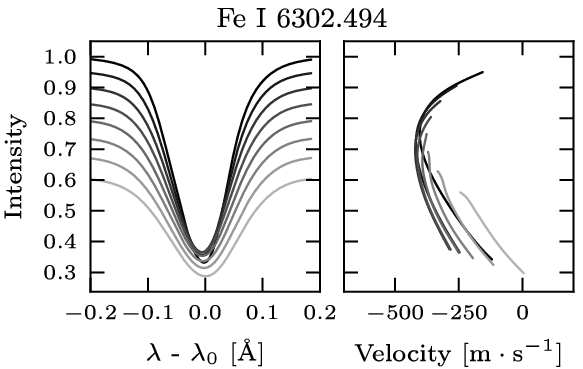}\includegraphics[]{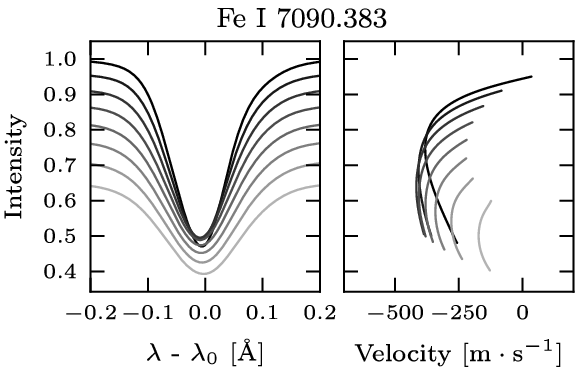}\includegraphics[]{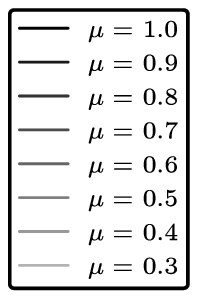}}\\
     	 \resizebox{\hsize}{!}{\includegraphics[]{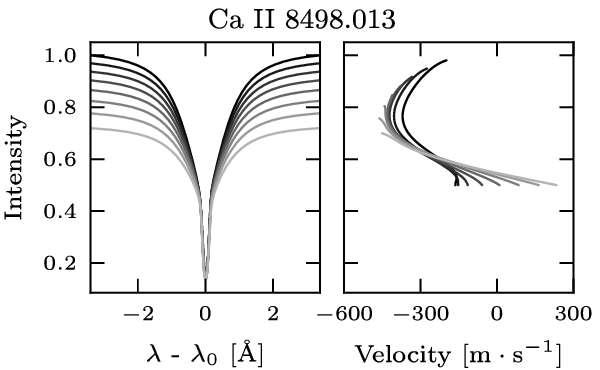}\includegraphics[]{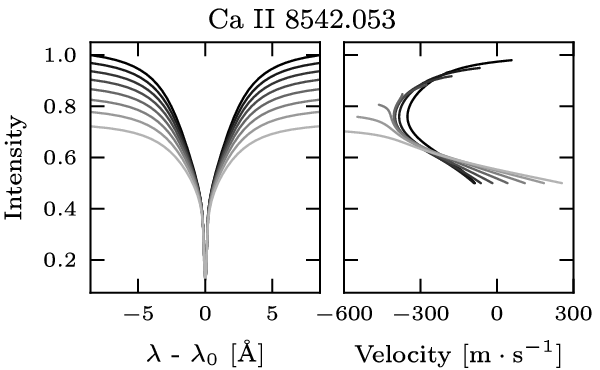}\includegraphics[]{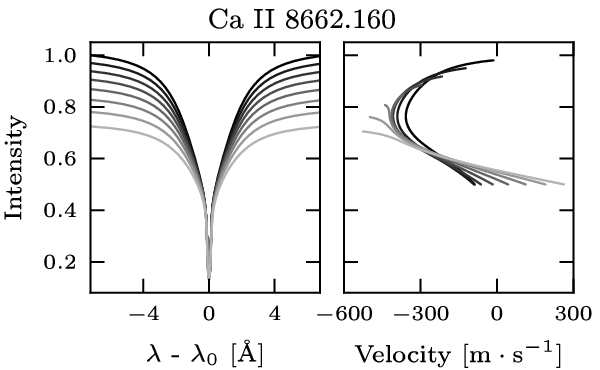}}
        \caption{Synthetic profiles and the corresponding bisectors resulting from the calculations of \ion{C}{i}, \ion{Fe}{i} and \ion{Ca}{ii} lines. The gray scale indicates the variation in the heliocentric angle from $\mu=1.0$ to $\mu=0.3$. The  bisectors for the \ion{Ca}{ii} lines are not shown in the core part.}
        \label{velpan}
\end{figure*}

In Fig.~\ref{velpan} the resulting line profiles for all heliocentric angles are plotted together, along with their bisectors. The \ion{C}{i}~5380~\AA\ line shows the strongest convective blueshift, an expected result given that the line is formed very deep due to the hight excitation potential of the transition. In deep layers, vertical velocities are high and the surface of the intergranular lanes becomes smaller. The blueshift decreases monotonously towards the limb. 

A variety of bisector shapes and $\mu$ dependences are found in the \ion{Fe}{i} lines. At $\mu=1$, the lines have the expected blueshifted convective C-shaped bisectors. When going off center, the blueshift for many lines increases, while the blueshift decreases again at even higher heliocentric angles, sometimes turning into a redshift, and the bisectors become more \textbackslash-shaped. Each line shows its own individual behavior -- function of its strength, excitation potential, and other properties. The line cores have been excluded from the bisectors for the \ion{Ca}{ii} lines. These wing bisectors show a progression from a C-shape to a \textbackslash-shape from center to limb. 

Table~\ref{tabmean} lists numerical values for the convective shifts computed as the mean of the bisector -- excluding the part closest to the continuum -- for each line and $\mu$ value, and Table~\ref{tabcores}  lists the line core velocities found when convolving the synthetic line profiles with the appropriate CRISP instrumental profile.

\subsection{Effect of spectral resolution}
The asymmetry of the lines will decrease with decreased spectral resolution. When comparing with observations, or using the theoretical profiles for velocity calibration, the profiles must be convolved to the proper spectral resolution. Figure~\ref{spectral_res} illustrates the dependence of the mean bisector, the center-of-gravity (COG) velocity, and the line-core velocity on spectral resolution. The core velocity is naturally the one most sensitive to spectral resolution. COG is least sensitive in this  specific case, but the deviations in all cases start at such spectral resolutions that we are content to use the mean bisector velocity in all our illustrations here. We do think that any practical use of our line profiles should use the detailed profile with the application of the appropriate instrumental smearing and weighting.

\subsection{The limb effect}\label{limbeff}
The bisectors change with heliocentric angle, and close to the limb the blueshift usually gets small and is sometimes turned into a redshift for the line cores. This limb effect \citep{1976MNRAS.177..687A, 1978SoPh...58..243B,1985SoPh...99...31B} was subject to much speculation before granulation was understood \citep[see][]{1982ARA&A..20...61D}. It seems that a comprehensive explanation is still lacking, which could be because the complicated nature of the situation precludes simple explanations. Nevertheless, we want to discuss it here in the light of our results.

The convective blueshift and line asymmetry at disk center are caused by the asymmetry of large bright granules with a relatively slow upward flow producing blueshifted lines and narrow dark intergranular lanes with fast downward flows producing redshifted lines. As we leave the disk center, a number of effects will change the shifts and the line profiles. Towards the limb, higher photospheric layers are sampled that have different velocity patterns and different statistical relations between intensity and line-of-sight velocities. Of particular interest is the pattern of inverse granulation \citep[e.g.,][]{2007A&A...461.1163C} that characterizes the photospheric structure above a certain height.   Also, horizontal velocities become more important as the viewing angle decreases and will soon dominate the contribution from vertical convective motions. There are also 3D effects due to the corrugated nature of the optical depth surface.  We can consider the following mechanisms that can decrease the blueshift and ultimately lead to a redshift in the average line profiles:
   \begin{figure}[]
      \centering
     	\resizebox{\hsize}{!}{\includegraphics[]{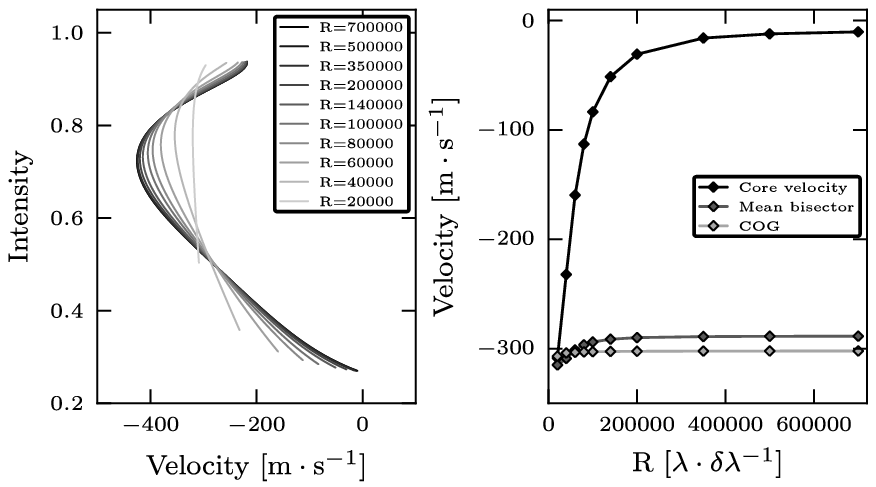}}
        \caption{\emph{Left:} Bisectors of the synthetic \ion{Fe}{i}~6301.5~\AA \ line as a function of the spectral resolution. \emph{Right:} Velocities measured in different ways from the profiles on the left panel, as a function of spectral resolution.}
        \label{spectral_res}
    \end{figure}

   \begin{figure}[]
      \centering
     	\resizebox{\hsize}{!}{\includegraphics[]{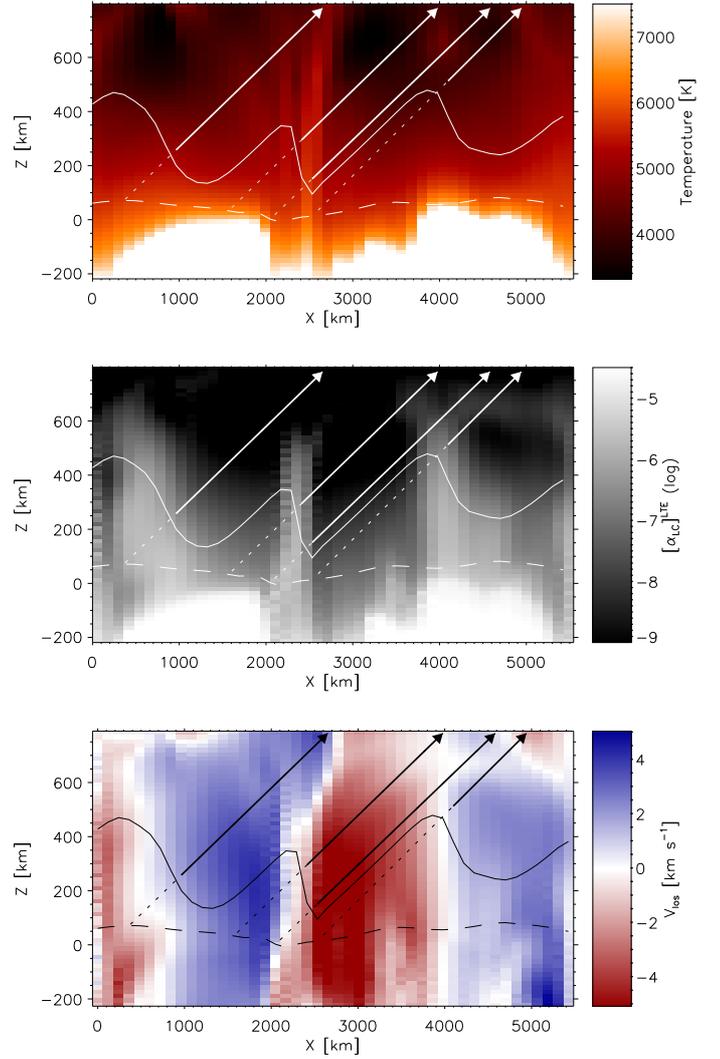}}
        \caption{Properties of the model along a slice of one of the snapshots at $\mu=0.3$: temperature, absorption coefficient at the core of \ion{Fe}{i}~6301.5~\AA, and line of sight velocity from top to bottom. The solid line that is overplotted over the three panels represents the layer where $\tau_{LC}=1$ along the line of sight. Four rays are shown as solid lines above $\tau_{LC}=1$ and dashed below this height. The bottom panel shows the line-of-sight velocities along the direction shown by the rays. The horizontal dashed line represents the layer where $\tau_{500}$ equals 1 at $\mu = 0.3$.}
        \label{layers}
   \end{figure}
   \begin{figure}[]
      \centering
	\resizebox{\hsize}{!}{\includegraphics[]{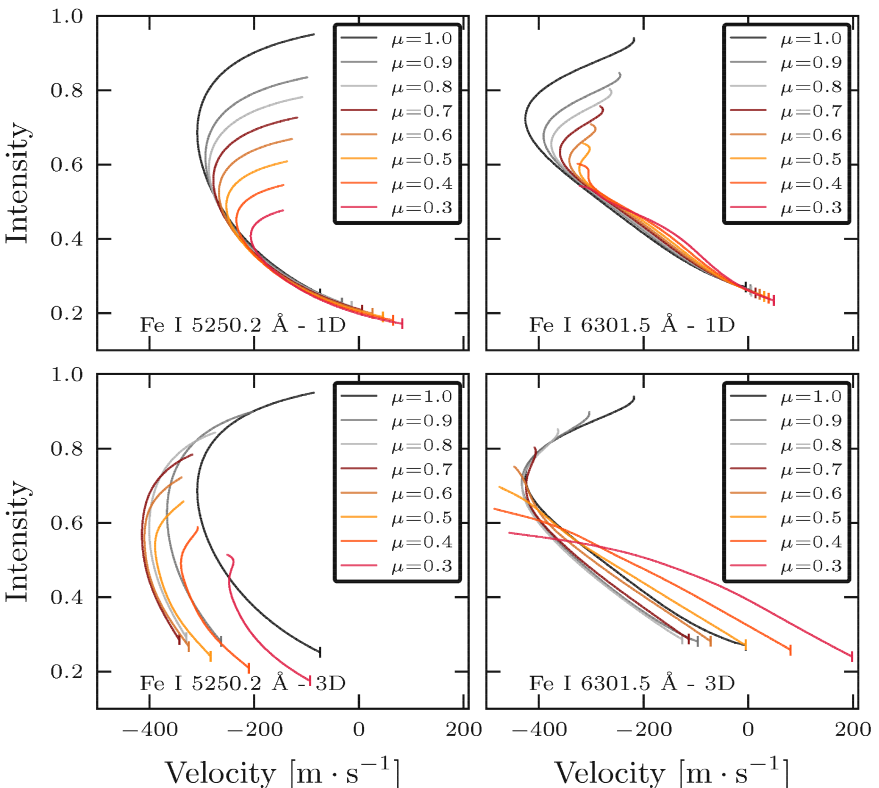}}
        \caption{Bisectors for two lines computed in the plane-parallel treatment, ``1D'', compared to the full ``3D'' result.}
        \label{1dtest}
    \end{figure}
    
\begin{enumerate}
\item\label{eff-ivhi} The intensity-velocity correlation changes as higher and higher levels in the photosphere are sampled by the line \citep[e.g.][]{2006ASPC..354...49J,2007A&A...461.1163C}. This is the same effect that causes the C-shape in disk-center bisectors.
\item\label{eff-ivp} The redshifted part of the granule is seen against a brighter background than the blueshifted part \citep{1978SoPh...58..243B,2000A&A...359..729A}.
\item\label{eff-corr} Near the limb, surface corrugation causes redshifted parts of granules to show larger areas to the observer \citep{1985SoPh...99...31B}. 
\item\label{eff-shield} Reversed granulation ``hills'' over intergranular lanes shield blueshifted parts of granules when observed  off disk center.
\end{enumerate}

Figure~\ref{layers} shows a cross section through a simulation snapshot illustrating line formation at $\mu=0.3$. Inspection of the simulation snapshots in this way can give support to all the listed mechanisms and also inspire invention of new ones. The effect of the hot wall over an intergranular lane (at $x\sim 2200$~km) is, for example, obvious. It blocks the view towards the blueshifted side of a granule. But a similar structure is also present over the center of a granule ($x\sim 4000$~km), seemingly blocking a redshifted side of a granule. One can thus argue that some of the mentioned mechanisms and others can produce a {\em blueshift} with decreasing $\mu$. Deciding which mechanisms  will dominate requires a quantitative analysis. We made a first effort in that direction.
   
Effects \ref{eff-ivp}, \ref{eff-corr}, and \ref{eff-shield} are true 3D effects that break the symmetry between redshifted limbside granular parts and blueshifted centerside parts. They should go away in a multi-1D treatment where each column in the simulation is treated as a plane-parallel atmosphere. The results of such a numerical experiments for two of our lines can be seen in Fig.~\ref{1dtest}. It is clear that for each line, the variation of bisectors with $\mu$ is smaller than in 3D. It is also monotonous and always in the sense that a profile becomes redder at the core of the line with decreasing $\mu$. The close following of the successive bisectors to that of $\mu=1$ is an illustration of the Eddington-Barbier relations: the $\mu$ dependence is equivalent to the depth dependence. We take this to mean that much of the center-to-limb variation of spectral line shifts is due to 3D effects. The latter either shifts the  bisectors to redder or bluer wavelengths, relative to the 1D case. For example, the blueshift relative to disk center that is seen in many lines for moderate heliocentric angles must be due to a 3D mechanism. We hypothesise that this comes from the screening of narrow downdrafts from view as the line of sight departs from the vertical.

\begin{table*}
\caption{\label{tabmean} Convective shifts in $\mbox{m}\ \mbox{s}^{-1}$ of the studied lines from $\mu = 1.0$ to $\mu = 0.3$. }
        \centering          
\begin{tabular}{c|c|cccccccc}
	\hline\hline
  Element & $\lambda_0$ (\AA) & $\mu=1.0$ & $\mu=0.9$ & $\mu=0.8$ & $\mu=0.7$ & $\mu=0.6$ & $\mu=0.5$ & $\mu=0.4$ & $\mu=0.3$ \\
       \hline  
\ion{C}{i}  &5380.34 &-948 & -936 & -922 & -910 & -897 & -888 & -880 & -844 \\
\ion{Fe}{i} &5250.21 &-294 & -372 & -410 & -426 & -420 & -394 & -346 & -259 \\
\ion{Fe}{i} &5250.65 &-304 & -354 & -370 & -369 & -348 & -308 & -245 & -140 \\
\ion{Fe}{i} &5576.09 &-334 & -378 & -388 & -383 & -360 & -317 & -249 & -140 \\
\ion{Fe}{i} &6082.71 &-441 & -495 & -522 & -536 & -531 & -511 & -473 & -397 \\
\ion{Fe}{i} &6301.50 &-307 & -349 & -358 & -351 & -325 & -279 & -211 & -101 \\
\ion{Fe}{i} &6302.49 &-335 & -387 & -404 & -403 & -381 & -342 & -280 & -176 \\
\ion{Fe}{i} &7090.38 &-352 & -396 & -409 & -407 & -389 & -354 & -302 & -210 \\
\ion{Ca}{ii}&8498.01 &-265 & -288 & -289 & -275 & -248 & -206 & -148 & -70  \\
\ion{Ca}{ii}&8542.05 &-335 & -364 & -374 & -372 & -359 & -337 & -307 & -269 \\
\ion{Ca}{ii}&8662.16 &-272 & -294 & -293 & -279 & -251 & -208 & -149 & -68  \\
\hline
\end{tabular}
\tablefoot{The bisector of each line was integrated from $95\%$ of the continuum intensity to the core of the line. In the case of the \ion{Ca}{ii} lines, the non-LTE core was excluded. For \ion{C}{i} the upper threshold was higher ($97 \%$) because the line is weak.}
\end{table*}

\begin{table*}
\caption{\label{tabcores} Line core velocities of the synthetic spectral lines (excluding \ion{Ca}{ii}) convolved with the appropriate theoretical profile of CRISP for each wavelength. }
        \centering
\begin{tabular}{c|c|cccccccc}
\hline \hline
  Element & $\lambda_0$ (\AA) & $\mu=1.0$ & $\mu=0.9$ & $\mu=0.8$ & $\mu=0.7$ & $\mu=0.6$ & $\mu=0.5$ & $\mu=0.4$ & $\mu=0.3$ \\
       \hline  
\ion{C}{I}  & 5380.34  & -928  & -895  & -867  & -840  & -820  & -810  & -795  & -764 \\
\ion{Fe}{I} & 5250.21  & -178  & -306  & -358  & -370  & -354  & -315  & -246  & -134 \\
\ion{Fe}{I} & 5250.65  & -115  & -196  & -221  & -213  & -177  & -115  & -26   & 108  \\
\ion{Fe}{I} & 5576.09  & -163  & -222  & -231  & -209  & -164  & -96   & -3    & 126  \\
\ion{Fe}{I} & 6082.71  & -403  & -470  & -499  & -508  & -500  & -480  & -433  & -341  \\
\ion{Fe}{I} & 6301.50  & -188  & -238  & -241  & -217  & -173  & -109  & -17   & 108  \\
\ion{Fe}{I} & 6302.49  & -255  & -326  & -347  & -338  & -306  & -255  & -178  & -57  \\
\ion{Fe}{I} & 7090.38  & -315  & -378  & -396  & -388  & -362  & -321  & -255  & -148 \\
\hline
\end{tabular}
\end{table*}

\subsection{Error analysis}
  \subsubsection{Non-LTE effects in \ion{Fe}{i} lines}\label{non-LTEe}
Our radiative transfer calculations assume LTE. The errors introduced by this approximation will to some extent be reduced by fitting the line profiles using an individual abundance value for each line as a free parameter. \citet{2001ApJ...550..970S} studied the impact of non-LTE effects on \ion{Fe}{i} and \ion{Fe}{ii} lines using a 3D simulation (but a 1D treatment of the non-LTE problem, thus 1.5D) and reported larger departures from LTE in granules than in intergranular lanes. Based on this work, we can expect the following two non-LTE mechanisms to dominate
\begin{enumerate}
	\item \emph{Overionization}: A line from a neutral species typically has lower line opacity than its LTE value because of overionization that is caused by the UV radiation field being stronger than the local Planck function. This causes the line to be formed deeper in the atmosphere and thus be weaker than in LTE.
	\item The non-LTE line source function $S^l_\nu$ can be expressed in terms of the departure coefficients of the populations from the upper ($b_j$) and lower ($b_i$) levels \citep[see][]{rutten-lectures} if stimulated emission is neglected,

\begin{equation}
	\centering
	S^l_\nu \sim \frac{b_j}{b_i} B_\nu(T)
	\label{snon-LTE}
\end{equation}
where $B_\nu(T)$ is the Planck function.
\emph{Photon losses} in the line itself lead to a decrease in $b_j$ and increase on $b_i$, producing a line source function that is lower than the Planck function. This leads to a stronger line than in LTE.	
\end{enumerate}

\begin{figure}[]
      \centering
      \resizebox{\hsize}{!}{\includegraphics[]{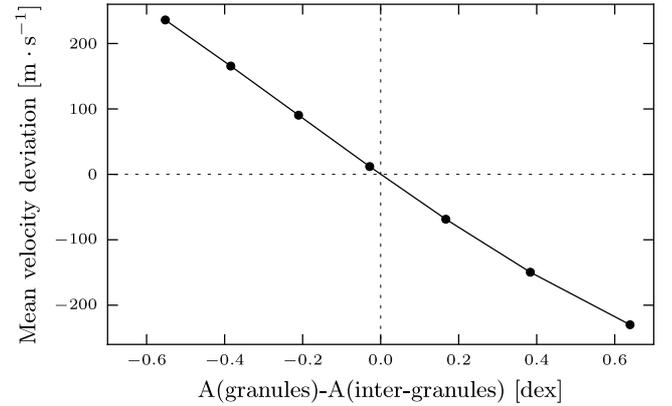}}
      \caption{Differential Doppler shift of the \ion{Fe}{i}~6082.7~\AA \ line produced by changes in the relative abundance between granules and inter-granules. The combinations of granular/intergranular abundances were fitted to reproduce the equivalenth width of the FTS atlas.} 
        \label{non-LTE_test}
\end{figure}

The first effect tends to be the most important for weak lines, while the second is found for strong lines because the line opacity dominates the background opacity. In some cases, the non-LTE and LTE profiles are very similar because the two effects cancel out. However, the presence of larger LTE departures in the granules than in the intergranular lanes not only changes the strength of the line, but also the convective line shift. We tested the sensitivity of the resulting line shift to such differential departures from LTE. One of the 3D snapshots was divided in granules and intergranular lanes based on the sign of the vertical velocities at $\tau_{500}=1$ and synthetic spectra were computed for different values of the iron abundance. The aim was to find the combinations of abundances in granules and intergranules that reproduce a spatially-averaged spectrum with the same equivalent width as the observed profile. We used a nonlinear least-squares fit, described in \citet{mpfit}, in order to infer these pairs of values. Figure~\ref{non-LTE_test} illustrates the average shift of the line as a function of abundance difference between the granules and the intergranular lanes. The reference was set where the abundance derived from the granules and intergranules is the same. \citet{2001ApJ...550..970S} find an average non-LTE effect on derived Fe abundances of 0.1 dex. If this number is taken as also typical of the difference between granule and intergranule abundance in Fig.~\ref{non-LTE_test}, the errors in the computed lineshift due to non-LTE effects should be smaller than $\pm 30 \ \mbox{m}\ \mbox{s}^{-1}$.
  \subsubsection{Profile fitting}
  
Due to the effects described in Section \ref{non-LTEe} and becuase the 3D models are only realistic up to a certain height, the resulting profiles show differences compared to the FTS atlas. The solar abundances inferred by \cite{2005ASPC..336...25A} normally lead to different results for the different lines, some presenting more damped wings than the atlas and others narrower ones.  The abundance is used as a free parameter in order to match the wings of the synthetic profiles for $\mu=1$ with the FTS atlas. The inner core is not included in the fitting procedure because generally non-LTE effects are expected to be larger there \citep[see][]{1982A&A...115..104R}, and the model snapshots become less reliable close to the upper boundary \citep[see][]{2000A&A...359..729A}.

\begin{figure}[]
      \centering
      \begin{center}
            \resizebox{\hsize}{!}{\includegraphics[]{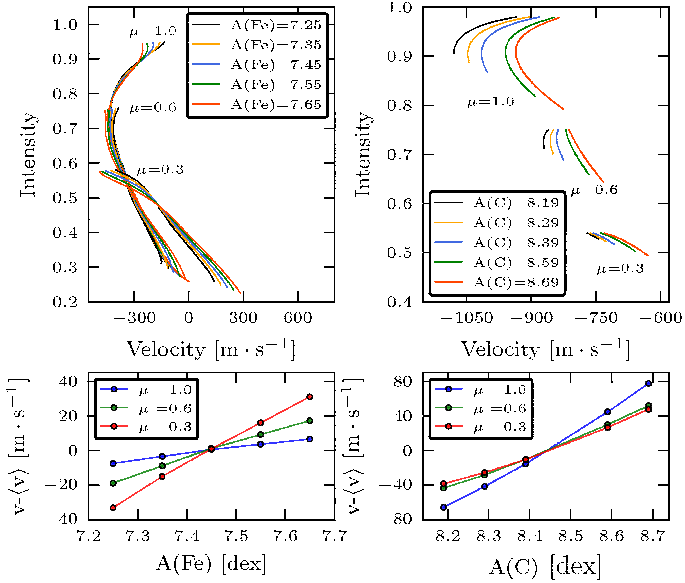}}
        \caption{Bisectors calculated for at $\mu=1.0,0.6,0.3$ for a range of abundances. The results are only from five snapshots which means that they cannot be directly compared with, e.g., Fig\,\ref{velpan}. \emph{Left:} abundance test for \ion{Fe}{i}~6301~\AA \ with values of the Fe abundance within [7.25-7.65] . \emph{Right:} test for \ion{C}{i}~5380~\AA, the values of the abundance span an unrealistic range compared to the nominal value [8.19-8.69].} 
          
        \label{abundtf}
      \end{center}
\end{figure}

As an indicator of the errors due to this fitting procedure, we performed test calculations to get the sensitivity of the bisectors to changes in the assumed abundance. The results are shown in Fig.~\ref{abundtf}. The left panels correspond to the \ion{Fe}{i}~6301~\AA \ line, calculated on three different heliocentric angles, for different values of the Fe abundance. The panels on the right correspond to the \ion{C}{i}~5380~\AA, where the abundance covers an unrealistic range of values $[8.19,8.69]$. This weak line, which is formed very close to the continuum, is more sensitive to changes in the abundance than  the  stronger \ion{Fe}{i} line. The lower panels show the relative velocity shift integrated over the whole bisector. We estimate that the uncertainties from the fitting procedure are within 0.1 dex in abundance which then corresponds to $\pm 10\ \mbox{m}\ \mbox{s}^{-1}$ for the integrated bisector.
\begin{figure*}[]
      \centering
      \begin{center}
            \resizebox{\hsize}{!}{\includegraphics[]{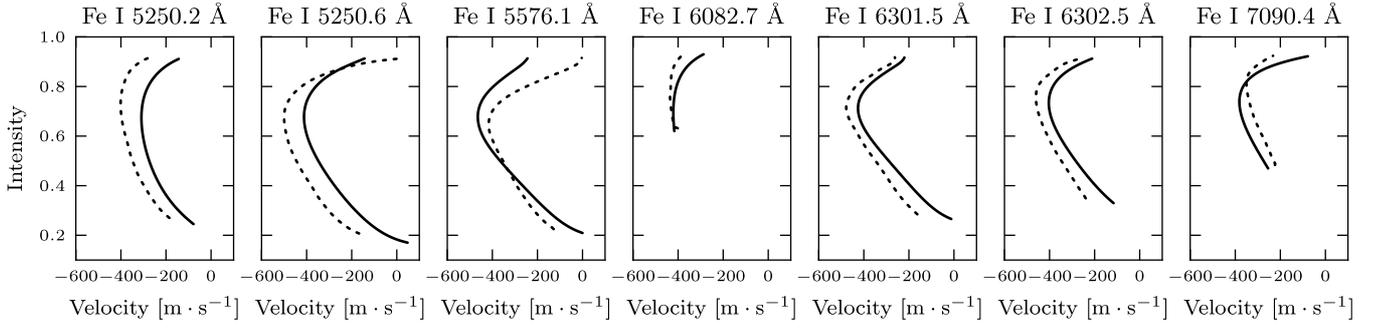}}
        \caption{\ion{Fe}{i} synthetic bisectors (fulldrawn) compared with bisectors measured from the FTS atlas (dashed).} 
        \label{ftscomp}
      \end{center}
\end{figure*}
  \subsubsection{Discussion}
From the tests above, the estimated error in the shift of the lines is less than $50\ \mbox{m}\ \mbox{s}^{-1}$. This means that a full non-LTE treatment combined with the exact atomic data should not change these results by more than this amount. The estimate does not, however, include the intrinsic errors of the simulation.

To check the realism of the simulations, we compared the synthetic line profiles at $\mu=1.0$ with the FTS atlas after applying a gravitational redshift on the former. The average synthetic \ion{C}{i} bisector lies 530~$\mbox{m}\ \mbox{s}^{-1}$ redwards of that measured from the FTS atlas, something that we deem mostly to be due to error in the laboratory wavelength of the line. At face value, our simulations implies a wavelength of 5380.3275~\AA\ on the FTS-atlas wavelength scale. This is within the cited uncertainty (0.02 \AA) of the laboratory value of 5380.3370 \AA\ from \cite{1966johansson}, which was used here. It is also comfortingly close to the 5380.3262 \AA\ found in a similar way by \citet{2007ApJ...655..615L} from their observations and simulations of magnetic regions. The bisectors for the \ion{Fe}{i} lines are displayed in Fig.~\ref{ftscomp}. The synthetic bisector shapes show an acceptable correspondence with those measured from the FTS atlas. The standard deviation of the differences in mean velocity is 56~$\mbox{m}\ \mbox{s}^{-1}$ which corresponds to approximately 1 m\AA. This is comparable to our estimate of the internal errors from non-LTE effects and fitting procedures, but once again there is strong reason to believe that a significant part of these deviations are due to errors in the adopted laboratory wavelengths. 

Finally we note that \citet{1998A&AS..131..431A} estimate the precision in FTS wavelengths to be on the order of 50~$\mbox{m}\ \mbox{s}^{-1}$. It seems likely that several of the error estimates discussed above are too conservative, but the synthetic profiles seem to pass the test and they should be correct within 50~$\mbox{m}\ \mbox{s}^{-1}$.
Observational checks of the off-center profiles are more difficult but would be welcome. Since photospheric modeling often becomes more uncertain for low $\mu$ values, it is natural to suspect that the accuracy of our convective line shifts will decrease towards the limb. 

The hydrodynamic granulation simulations do not include magnetic fields. This is potentially the most important source of errors in the practical application of these synthetic line profiles for velocity calibrations. For example, \citet{2007ApJ...655..615L} estimated a difference of 200~$\mbox{m}\ \mbox{s}^{-1}$ in convective shift for the \ion{C}{i}~5380~\AA\ line between non-magnetic and active granulation. (This was mostly caused by the changing intensity and velocity pattern and not by Zeeman splitting.) Any calibration work must use the least magnetic photospheric region available. However, how strong the quiet Sun magnetic fields are is still an ongoing debate \citep[e.g.][]{2008A&A...477..953M, 2009ApJ...693.1728P, 2009SSRv..144..275D}.
\subsection{Calibration of SST/CRISP observations}
A quiet Sun dataset acquired with SST/CRISP at $\mu=1.0$ was calibrated using the current results. The \ion{Fe}{i}~6302~\AA\ line was sampled with eight wavelength points, evenly distributed with a separation of $\mbox{d}\lambda=48$ m\AA \ and  processed using the image restoration code \textsc{MOMFBD} \citep{2005SoPh..228..191V}. The telecentric optical design of \textsc{CRISP} \citep{2006A&A...447.1111S} is optimum for high image quality, but it introduces instrumental field-dependent wavelength shifts produced by irregularities on the surface of the etalons, which are commonly known as cavity errors. The latter complicate the task of obtaining the spatially-averaged spectrum, as the profile is effectively broadened by these random pixel-to-pixel shifts. The average spectrum was computed with a least-squares-fit of all the data points to a spline which was characterized with ten parameters. For this purpose, all the individual spectra were corrected for cavity errors and then arranged in a large array, which was sorted in increasing wavelength. 

\begin{figure}[]
      \centering
        \resizebox{\hsize}{!}{\includegraphics[]{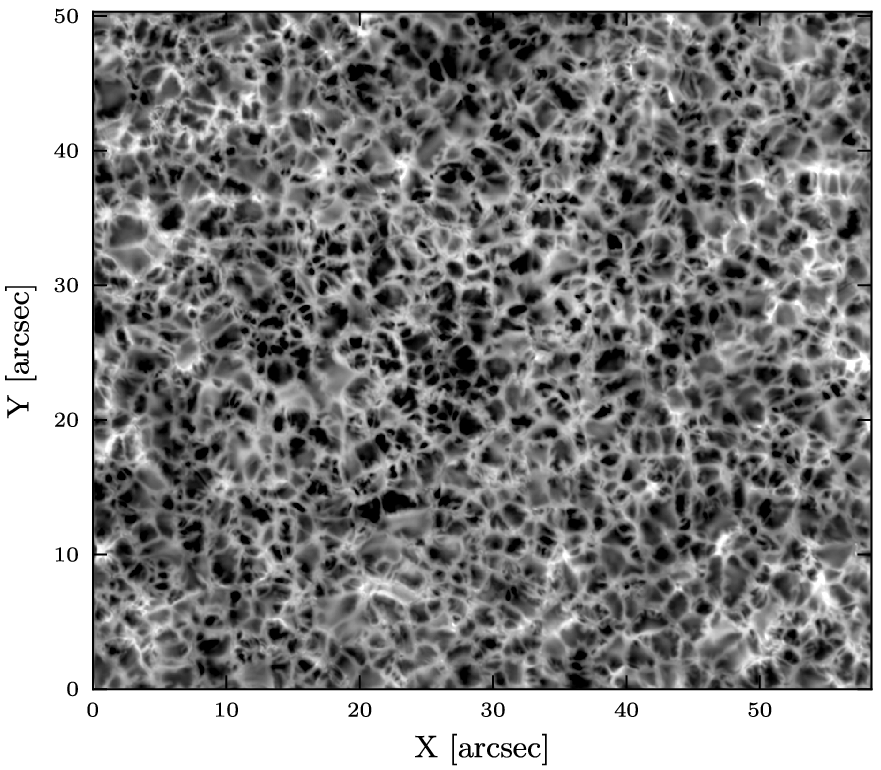}}
     	\resizebox{\hsize}{!}{\includegraphics[]{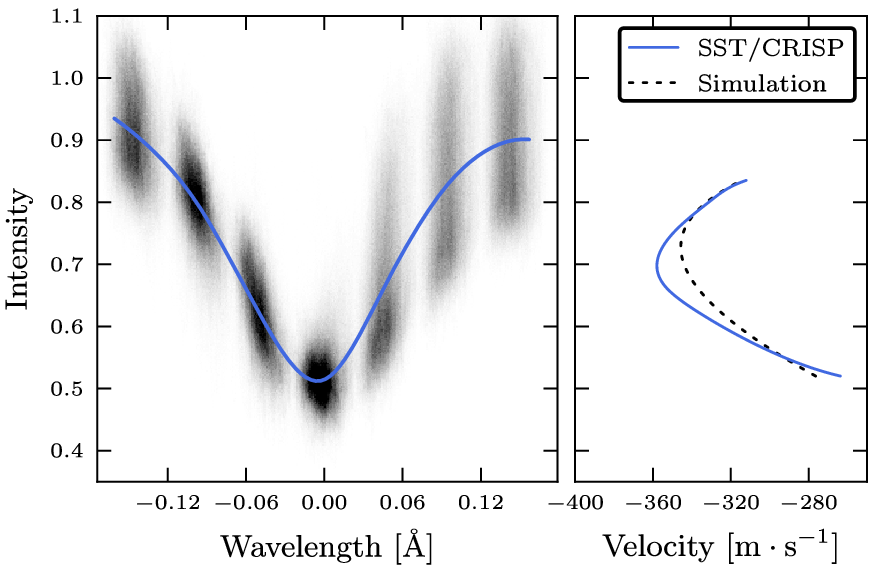}}
        \caption{\emph{Top:} Doppler map of quiet sun granulation observed with SST/CRISP at \ion{Fe}{i}~6302~\AA. The image gray scale is clipped to $\pm1500 \ \mbox{m}\ \mbox{s}^{-1}$. \emph{Bottom-left:} Spatially-averaged profile computed along the image shown in the top panel. The density map represents the spatially-resolved profiles corrected for instrumental cavity errors. The average profile was computed using a least squares fit of all the points to a spline, consisting of 10 parameters.  \emph{Bottom-right:} Bisector of the observed spatially-averaged profile. The velocity calibration was carried out using our synthetic profile convolved with the CRISP instrumental profile.}
        \label{crisp_prof}
    \end{figure}

To carry out the velocity calibration, the synthetic spectrum obtained from the 3D simulation was convolved with the theoretical instrumental profile of CRISP, which included the effects of the beam converging at $f/165$.  The velocity calibration was determined by finding the constant wavelength shift to the observed bisector that minimized the difference to the synthetic one. Figure~\ref{crisp_prof} illustrates the results of the calibration, where  the Doppler map at the core of the line is scaled between $\pm 1500 \ \mbox{m}\ \mbox{s}^{-1}$. The bottom-left panel contains a density map of the spatially resolved spectra, corrected for cavity errors. The dispersion of points around the seven nominal wavelengths appears after correcting for the cavity errors in combination with the granulation contrast. The bisector of our calibrated data is overplotted to the bisector of the synthetic spectrum that was convolved with the instrumental profile of CRISP. There is a telluric blend on the red wing of the observed profile, not present in the synthetic spectrum, that  pollutes the upper part of the bisector.
\section{Usage of electronic data}\label{edata}
The synthetic profiles that have been presented in this study can be accessed electronically. The spectral response of any instrument produces broadening and can introduce asymmetries in the observed profiles. Thus it is \emph{imperative} that the profiles be degraded with the proper instrumental response before they are used to calibrate observations.

The quiet Sun's average spectrum can be computed from quiet regions in the observations. However, the presence of p-modes (oscillations) will produce an unrealistic shift in the observed average profile. Thus, it should be averaged in time and space, or at least from a sufficiently large sample of quiet Sun spectra.

We have been cautious in the treatment of line cores, but the comparison with the FTS atlas suggests that the errors in the cores are small. Except for the \ion{Ca}{ii} IR triplet, we do not discourage including the cores in velocity calibrations. We think, however, that a velocity measure that weigh in more of the profile will give a more reliable result.
\section{Conclusions}\label{conclusions}
We made use of the approach described by \citep{2007ApJ...655..615L} to obtain an absolute reference for velocities in solar observations. For this purpose, we calibrated the shift in synthetic spectra produced by convective motions at different heliocentric angles for \ion{C}{i}, \ion{Fe}{i}, and \ion{Ca}{ii} lines as commonly used for solar physics diagnostics. We propose to use the profiles as absolute local references for line-of-sight velocities to allow accurate velocity measurements.

A realistic 3D hydrodynamical simulation of the non-magnetic solar photosphere was used to synthesize the spectra. The computations were done in LTE for \ion{C}{i} and \ion{Fe}{i} lines whereas non-LTE computations were carried out in the \ion{Ca}{ii} infrared lines. As the conversion from wavelength to velocity scale is done using the same laboratory wavelength as used for computing the profiles, the measured line shifts are relatively unaffected by errors in the central wavelength. Uncertainties in the rest of atomic parameters are partially compensated when the profiles are fitted to reproduce the strength of the FTS atlas.

Our results show a systematic center-to-limb variation of the bisectors, which is determined to a large extent by the 3D structure of the atmosphere, which is projected along the line-of-sight. However, there is a small amount of redshift (relative to $\mu=1.0$) produced by the change in the formation height with the heliocentric angle (non-3D effects), which is not related to projection effects. A careful analysis of the models suggests that a large number of 3D snapshots must be used in studies of line shifts so the oscillations are properly sampled and the statistics of granules and intergranules are complete, whereas non-LTE effects seem to have a minor influence on the shift of our \ion{Fe}{i} lines. The largest uncertainty is related to the non-magnetic nature of the simulation. Nevertheless, we estimate our calibrations to be accurate up to $\sim50 \ \mbox{m}\ \mbox{s}^{-1}$  at $\mu=1$, remaining untested towards the solar limb. 

We would like to emphasize the somewhat limited usability of the \ion{Ca}{ii} calibration, as observations are polluted by a significant number of blends, which should be masked in order to compare with the calibration data.

Our synthetic profiles reproduce spatially-averaged solar observations accurately, but it would be interesting to test for the impact of magnetic fields on the convective shift of the lines, and confirm the reliability of our calculations near the limb.
 
\begin{acknowledgements}
We thank Tiago Pereira for providing the 3D snapshots of the HD simulation and Roald Schnerr and Michiel van Noort for providing reduced data from the SST/CRISP. We are most grateful to Paul Barklem for providing unpublished cross-sections for collisional broadening. This research project has been supported by a Marie Curie Early Stage Research Training Fellowship of the European Community's Sixth Framework Program under contract number MEST-CT-2005-020395: The USO-SP International School for Solar Physics. The Swedish 1-m Solar Telescope is operated on the island of La Palma by the Institute for Solar Physics of the Royal Swedish Academy of Sciences in the Spanish Observatorio del Roque de los Muchachos of the Instituto de Astrof\'isica de Canarias.
\end{acknowledgements}

\bibliographystyle{aa}
\bibliography{15664}
\end{document}